\documentstyle[12pt]{article}
\begin{document}

\title {ON THE ROLE OF DENSITY INHOMOGENEITY AND
LOCAL ANISOTROPY IN THE FATE OF SPHERICAL COLLAPSE}
\author{L. Herrera\thanks{Also at Centro de
Astrof\'\i sica Te\'orica, M\'erida, Venezuela and Departamento de F\'\i
sica, Universidad Central de Venezuela, Caracas, Venezuela.}\ ,
A. Di Prisco\thanks{On leave from Departamento de F\'\i sica, Universidad
Central de Venezuela, Caracas, Venezuela.}\ ,
J. L. Hern\'andez-Pastora \\
\'Area de F\'\i sica Te\'orica\\
Facultad de Ciencias\\
Universidad de Salamanca\\
37008, Salamanca, Espa\~na.\\
and
\and
N. O. Santos\\
Departamento de Astrof\'\i sica\\
CNPq-Observat\'orio Nacional\\
Rua General Jos\'e Cristino 77\\
20921-400 Rio de Janeiro-RJ, Brazil.}
\date{}
\maketitle

\begin{abstract}
We obtain an expression for the active gravitational mass
of a collapsing fluid distribution, which brings out the
role of density inhomogeneity and local anisotropy in
the fate of spherical collapse.
\end{abstract}

\section{Introduction}
It is well established that density inhomogeneities play
an important role in the collapse of a spherical dust
cloud. In particular they may lead to the formation of
naked singularities \cite{Var}, in contrast with the
homogeneous case leading to a black hole \cite{Opp}.
However, it is unclear what is the physical reason (if any)
behind the link between the final fate of collapse and the
presence (or absence) of density inhomogeneities.

\noindent
It is the purpose of this work to explore the ways in which
density inhomogeneities may affect the spherical collapse.
To do so, we shall obtain an expression for the active
gravitational mass (the Tolman mass) containing explicitly
a measure of density inhomogeneity (among other factors).
Since our expression is obtained for a general (locally
anisotropic) fluid, it also contains a term depending
on local anisotropy. We shall see that this term plays
a similar role to the density inhomogeneity term,
thereby suggesting  that local anisotropy might also lead
to the formation of naked singularities.

\noindent
As a by-product of this study we shall find some expressions
linking the Weyl tensor and the mass function  to the energy density
inhomogeneity and local
anisotropy.

\noindent
The paper is organized as follows. In the next section
the field equations and other useful formulae
are introduced.
In section 3 we introduce the Tolman mass and discuss
its physical meaning.
In Section 4 we derive an expression for the Tolman mass
and evaluate it at the moment the system departs
from hydrostatic equilibrium.
A discussion of this expression is presented in the last section.

\section{Field Equations and Conventions}

\noindent
We consider a spherically symmetric distribution of collapsing
fluid, which for completeness we assume to be anisotropic and
bounded by a
spherical surface $\Sigma$.
The line element is given in Schwarzschild-like coordinates by

\begin{equation}
ds^2=e^{\nu} dt^2 - e^{\lambda} dr^2 -
r^2 \left( d\theta^2 + sin^2\theta d\phi^2 \right)
\label{metric}
\end{equation}

\noindent
where $\nu$ and $\lambda$ are functions of $t$ and $r$ .
The coordinates are: $x^0=t; \, x^1=r; \, x^2=\theta; \, x^3=\phi$.

The metric (\ref{metric}) has to satisfy Einstein field equations

\begin{equation}
G^\mu_\nu=-8\pi T^\mu_\nu
\label{Efeq}
\end{equation}

\noindent
which in our case read \cite{Bon}:

\begin{equation}
-8\pi T^0_0=-\frac{1}{r^2}+e^{-\lambda}
\left(\frac{1}{r^2}-\frac{\lambda'}{r} \right)
\label{feq00}
\end{equation}

\begin{equation}
-8\pi T^1_1=-\frac{1}{r^2}+e^{-\lambda}
\left(\frac{1}{r^2}+\frac{\nu'}{r}\right)
\label{feq11}
\end{equation}

\begin{eqnarray}
-8\pi T^2_2  =  -  8\pi T^3_3 = & - &\frac{e^{-\nu}}{4}\left(2\ddot\lambda+
\dot\lambda(\dot\lambda-\dot\nu)\right) \nonumber \\
& + & \frac{e^{-\lambda}}{4}
\left(2\nu''+\nu'^2 -
\lambda'\nu' + 2\frac{\nu' - \lambda'}{r}\right)
\label{feq2233}
\end{eqnarray}

\begin{equation}
-8\pi T_{01}=-\frac{\dot\lambda}{r}
\label{feq01}
\end{equation}

\noindent
where dots and primes stand for partial differentiation with respect
to $t$ and $r$
respectively.

\noindent
In order to give physical significance to the $T^{\mu}_{\nu}$ components
we apply the Bondi approach \cite{Bon},
i.e we introduce    local Minkowski
coordinates ($\tau, x, y, z$), defined by

$$d\tau=e^{\nu/2}dt\,\qquad\,dx=e^{\lambda/2}dr\,\qquad\,
dy=rd\theta\,\qquad\, dz=rsin\theta d\phi$$

\noindent
Then, denoting the Minkowski components of the energy tensor by a bar,
we have

$$\bar T^0_0=T^0_0\,\qquad\,
\bar T^1_1=T^1_1\,\qquad\,\bar T^2_2=T^2_2\,\qquad\,
\bar T^3_3=T^3_3\,\qquad\,\bar T_{01}=e^{-(\nu+\lambda)/2}T_{01}$$

\noindent
Next we suppose that, when viewed by an observer moving relative to these
coordinates with velocity $\omega$ in the radial direction, the physical
content  of space consists of an anisotropic fluid of energy density $\rho$,
radial pressure $P_r$ and tangential pressure $P_\bot$.
Thus, when viewed by this moving observer, the covariant energy-momentum
tensor in
Minkowski coordinates is

\[ \left(\begin{array}{cccc}
\rho    &  0        &   0     &   0    \\
0       &  P_r      &   0     &   0    \\
0       &   0       & P_\bot  &   0    \\
0       &   0       &   0     &   P_\bot
\end{array} \right) \]

\noindent
Then a Lorentz transformation readily shows that

\begin{equation}
T^0_0=\bar T^0_0= \frac{\rho + P_r \omega^2 }{1 - \omega^2}
\label{T00}
\end{equation}

\begin{equation}
T^1_1=\bar T^1_1=-\frac{ P_r + \rho \omega^2}{1 - \omega^2}
\label{T11}
\end{equation}

\begin{equation}
T^2_2=T^3_3=\bar T^2_2=\bar T^3_3=-P_\bot
\label{T2233}
\end{equation}

\begin{equation}
T_{01}=e^{(\nu + \lambda)/2} \bar T_{01}=
-\frac{(\rho + P_r) \omega e^{(\nu + \lambda)/2}}{1 - \omega^2}
\label{T01}
\end{equation}

\noindent
Note that the velocity in the ($t,r,\theta,\phi$) system, $dr/dt$,
is related to $\omega$ by

\begin{equation}
\omega=\frac{dr}{dt}\,e^{(\lambda-\nu)/2}
\label{omega}
\end{equation}

\noindent
 Outside of the fluid , the spacetime is
Schwarzschild,

\begin{equation}
ds^2= \left(1-\frac{2M}{r}\right) dt^2
- \left(1-\frac{2M}{r}\right)^{-1} dr^2
- r^2 \left(d\theta^2 + sin^2\theta d\phi^2 \right)
\label{Sch}
\end{equation}

\noindent
In order to match the two metrics smoothly   on the boundary surface
$r=r_\Sigma(t)$, we  require  continuity of the first
and second fundamental
forms across that surface. As result of this matching we obtain
the well known result

\begin{equation}
\left[P_r\right]_\Sigma = 0
\label{Psig}
\end{equation}

\noindent
 The radial component of the
conservation law

\begin{equation}
T^\mu_{\nu;\mu}=0
\label{dTmn}
\end{equation}

\noindent
gives

\begin{equation}
\left(-8\pi T^1_1\right)'=\frac{16\pi}{r} \left(T^1_1-T^2_2\right)
+ 4\pi \nu' \left(T^1_1-T^0_0\right) +
\frac{e^{-\nu}}{r} \left(\ddot\lambda + \frac{\dot\lambda^2}{2}
- \frac{\dot\lambda \dot\nu}{2}\right)
\label{T1p}
\end{equation}

\noindent
which in the static case becomes

\begin{equation}
P'_r=-\frac{\nu'}{2}\left(\rho+P_r\right)+
\frac{2\left(P_\bot-P_r\right)}{r}
\label{Prp}
\end{equation}

\noindent
representing the generalization of the Tolman-Oppenheimer-Volkof equation
for anisotropic fluids \cite{BowLia}.

\noindent
For the next sections it will be useful to calculate the
components of the Weyl tensor. Using Maple V, it is found
 that all non-vanishing components are
proportional to

\begin{eqnarray}
W \equiv \frac{r}{2} C^{3}_{232} & = & W_{(s)} + \frac{r^3 e^{-\nu}}{12}
\left(\ddot\lambda + \frac{\dot\lambda^2}{2} -
\frac{\dot\lambda \dot\nu}{2}\right)
\label{W}
\end{eqnarray}

\noindent

where

\begin{equation}
W_{(s)} =
\frac{r^3 e^{-\lambda}}{6}
\left( \frac{e^\lambda}{r^2} - \frac{1}{r^2} +
\frac{\nu' \lambda'}{4} - \frac{\nu'^2}{4} -
\frac{\nu''}{2} - \frac{\lambda'}{2r} + \frac{\nu'}{2r} \right)
\label{Ws}
\end{equation}

\noindent
 corresponds
to the contribution in the static case.

\noindent
Finally, let us introduce the mass function, defined by \cite{MisSh,Cah}

\begin{equation}
m(r,t) = \frac{1}{2} r R^{3}_{232}
\label{fm}
\end{equation}

\noindent
where the Riemann component for metric (\ref{metric}) is given by

\begin{equation}
R^{3}_{232} = 1 - e^{-\lambda}
\label{R}
\end{equation}

\noindent
Then using the field equation (\ref{feq00}) we obtain
the well known expression

\begin{equation}
m(r,t) = 4 \pi \int^r_0{r^2 T^0_0 dr}
\label{m}
\end{equation}

\noindent
Now, considering the definition of the Weyl tensor

\begin{equation}
C^{\alpha}_{\beta \gamma \delta} = R^{\alpha}_{\beta \gamma \delta}
- \frac{1}{2} R^\alpha_\gamma g_{\beta \delta}
+ \frac{1}{2} R_{\beta \gamma} \delta^{\alpha}_{\delta}
- \frac{1}{2} R_{\beta \delta} \delta^{\alpha}_{\gamma}
+ \frac{1}{2} R^\alpha_\delta g_{\beta \gamma}
+ \frac{1}{6} R \left(\delta^{\alpha}_{\gamma} g_{\beta \delta} -
g_{\beta \gamma} \delta^{\alpha}_{\delta}\right)
\label{defC}
\end{equation}

\noindent
and eqs. (\ref{feq00}), (\ref{feq11}), (\ref{feq2233}),
(\ref{W}) and (\ref{fm}),  it follows that

\begin{equation}
m = \frac{4 \pi}{3} r^3 \left(T^0_0 + T^1_1 - T^2_2\right) + W
\label{mW}
\end{equation}

\noindent
Differentiating (\ref{mW}) with respect to r and using (\ref{m})
we get

\begin{equation}
W' = - \frac{4 \pi}{3} r^3 \left(T^0_0\right)' +
\frac{4 \pi}{3} \left[r^3 \left(T^2_2 - T^1_1\right)\right]'
\label{Wpri}
\end{equation}

\noindent
and integrating

\begin{equation}
W = - \frac{4 \pi}{3} \int^r_0{r^3 \left(T^0_0\right)' dr} +
\frac{4 \pi}{3} r^3 \left(T^2_2 - T^1_1\right)
\label{Wint}
\end{equation}

\noindent
Finally, inserting (\ref{Wint}) into (\ref{mW}) we obtain

\begin{equation}
m(r,t) = \frac{4 \pi}{3} r^3 T^0_0 -
\frac{4 \pi}{3} \int^r_0{r^3 \left(T^0_0\right)'dr}
\label{mT00}
\end{equation}

\noindent
Observe that expressions (\ref{mW}), (\ref{Wint}) and
(\ref{mT00}) are the same for the static (or slowly
evolving) case \cite{HeSa}. Of course, in the latter case
$T^0_0$ and $- T^1_1$ denote the proper energy density
and radial pressure, which is no longer true in the general
case as can be seen from eqs.(\ref{T00}) and (\ref{T11}).

\noindent
Instead of dealing with the mass function we shall now use the Tolman
mass, which, as will be seen in the next section, appears to embody
the physical idea of active gravitational mass better than
the mass function.

\section{The Tolman mass}

\noindent
The Tolman mass for a spherically symmetric distribution
of matter is given by (eq.(24) in \cite{To})

\begin{eqnarray}
m_T = & &  4 \pi \int^{r_\Sigma}_{0}{r^2 e^{(\nu+\lambda)/2}
\left(T^0_0 - T^1_1 - 2 T^2_2\right) dr}\nonumber \\
& + & \frac{1}{2} \int^{r_\Sigma}_{0}{r^2 e^{(\nu+\lambda)/2}
\frac{\partial}{\partial t}
\left(\frac{\partial L}{\partial \left[\partial
\left(g^{\alpha \beta} \sqrt{-g}\right) / \partial t\right]}\right)
g^{\alpha \beta}dr}
\label{Tol}
\end{eqnarray}

\noindent
where $L$ denotes the usual gravitational lagrangian density
(eq.(10) in \cite{To}). Although Tolman's formula was introduced
as a measure of the total energy of the system, with no commitment
to its localization, we shall define the mass within a sphere of
radius $r$, completely inside $\Sigma$, as

\begin{eqnarray}
m_T = & &  4 \pi \int^{r}_{0}{r^2 e^{(\nu+\lambda)/2}
\left(T^0_0 - T^1_1 - 2 T^2_2\right) dr}\nonumber \\
& + & \frac{1}{2} \int^{r}_{0}{r^2 e^{(\nu+\lambda)/2}
\frac{\partial}{\partial t}
\left(\frac{\partial L}{\partial \left[\partial
\left(g^{\alpha \beta} \sqrt{-g}\right) / \partial t\right]}\right)
g^{\alpha \beta}dr}
\label{Tolin}
\end{eqnarray}

\noindent
This extension of the global concept of energy to a local level
\cite{Coo} is suggested by the conspicuous role played by
$m_T$ as the ``effective gravitational mass'', which will be
exhibited below.
 Even though Tolman's definition is not
without its problems \cite{Coo,Deu}, we shall see that $m_T$,
as defined by (\ref{Tolin}), is a good measure of the
active gravitational mass, at least for the system under
consideration.

\noindent
Let us now evaluate expression (\ref{Tolin}). The first
integral in that expression

\begin{equation}
I \equiv 4 \pi \int^{r}_{0}{r^2 e^{(\nu+\lambda)/2}
\left(T^0_0 - T^1_1 - 2 T^2_2\right) dr}
\label{I}
\end{equation}

\noindent
may be transformed as follows. Integrating by parts and using
(\ref{m}), we obtain

\begin{eqnarray}
I = & & e^{(\nu+\lambda)/2} \left[m(r,t) - \frac{4 \pi}{3} r^3
\left(T^1_1 + 2 T^2_2\right)\right] \nonumber \\
& - & \int^r_0{e^{(\nu+\lambda)/2} \left(\frac{\nu' + \lambda'}{2}\right)
\left[m(r,t) - \frac{4 \pi}{3} r^3
\left(T^1_1 + 2 T^2_2\right)\right] dr} + \nonumber \\
& + & \int^r_0{\frac{4 \pi}{3} r^3 e^{(\nu+\lambda)/2} \left[
\left(T^1_1\right)' + 2 \left(T^2_2\right)'\right] dr}
\label{Ipart}
\end{eqnarray}

\noindent
 Using the field equations and (\ref{T1p}) we then obtain

\begin{eqnarray}
I = & & e^{(\nu+\lambda)/2} \left[m(r,t) - \frac{4 \pi}{3} r^3 T^1_1\right]
\nonumber \\
& - & \int^r_0{e^{(\lambda-\nu)/2} \frac{r^2}{2}
\left(\ddot\lambda + \frac{\dot\lambda^2}{2}
- \frac{\dot\lambda \dot\nu}{2}\right) dr}
\label{Ifin}
\end{eqnarray}

\noindent
 From
(eq.(13) in \cite{To})

\begin{equation}
\frac{\partial}{\partial t}
\left(\frac{\partial L}{\partial \left[\partial
\left(g^{\alpha \beta} \sqrt{-g}\right) / \partial t\right]}\right) =
- \Gamma^{0}_{\alpha \beta} + \frac{1}{2} \delta^0_\alpha
\Gamma^{\sigma}_{\beta \sigma}
+ \frac{1}{2} \delta^0_\beta
\Gamma^{\sigma}_{\alpha \sigma}
\label{13To}
\end{equation}

\noindent
and so the second integral $(II)$ in (\ref{Tolin}) may be expressed as

\begin{equation}
II = \frac{1}{2} \int^r_0{r^2 e^{(\lambda-\nu)/2}
\left(\ddot\lambda + \frac{\dot\lambda^2}{2} -
\frac{\dot\lambda \dot\nu}{2}\right)  dr}
\label{II}
\end{equation}

\noindent
Thus

\begin{equation}
m_T \equiv I + II = e^{(\nu + \lambda)/2}
\left[m(r,t) - 4 \pi r^3 T^1_1\right]
\label{I+II}
\end{equation}

\noindent
This is, formally, the same
expression for $m_T$ in terms of $m$ and $T^1_1$ that
appears in the static (or quasi-static) case
(eq.(25) in \cite{HeSa}).

\noindent
Replacing $T^1_1$ by (\ref{feq11}) and $m$ by (\ref{fm}) and (\ref{R}),
one  also finds

\begin{equation}
m_T = e^{(\nu - \lambda)/2} \, \nu' \, \frac{r^2}{2}
\label{mT}
\end{equation}

\noindent
This last equation brings out the physical meaning of $m_T$ as the
active gravitational mass. Indeed, it can be easily shown \cite{Gro}
that the gravitational acceleration ($a$) of a test particle,
instantaneously at rest in a static gravitational field, as measured
with standard rods and coordinate clock is given by

\begin{equation}
a = - \frac{e^{(\nu - \lambda)/2} \, \nu'}{2} = - \frac{m_T}{r^2}
\label{a}
\end{equation}

\noindent
A similar conclusion can be obtained by inspection of eq.(\ref{Prp})
(valid only in the static or quasi-static case) \cite{Lig}.
In fact, the first term on the right side of this equation
(the ``gravitational force'' term) is a product of the ``passive''
gravitational mass density $(\rho + P_r)$ and a term proportional
to $m_T/r^2$.

\noindent
In the next section we shall get another expression for $m_T$,
which appears to be more suitable for the treatment of
the problem under consideration. This expression
will be evaluated immediately after the system departs
from equilibrium, where ``immediately'' means on a
time-scale such that the value of $\omega$ remains unchanged.
Therefore the physical meaning of $m_T$, as the active
gravitational mass obtained for the static (and quasi-static)
case, may be safely extrapolated to the non-static case within
this time-scale.

\section{Tolman mass, density inhomogeneity and local anisotropy}

\noindent
The required expression for the Tolman mass will be obtained as follows.
Taking the $r$-derivative of (\ref{mT}) we obtain

\begin{equation}
r m'_T = \frac{e^{(\nu-\lambda)/2}}{2} \, r^3 \,
\left[\frac{\nu'^2}{2} - \frac{\lambda' \nu'}{2} + \nu'' +
\frac{2 \nu'}{r}\right]
\label{rde}
\end{equation}

\noindent
On the other hand, it follows from (\ref{I+II}) and (\ref{mW}) that

\begin{equation}
3 m_T = e^{(\nu + \lambda)} \left[4 \pi r^3
\left(T^0_0 - 2 T^1_1 - T^2_2\right) + 3 W\right]
\label{3m}
\end{equation}

\noindent
Combining (\ref{rde}) and (\ref{3m}) and using the
field equations and (\ref{Ws}), we obtain

\begin{eqnarray}
r m'_T - 3 m_T & = & e^{(\nu + \lambda)/2} \left[
4 \pi r^3 \left(T^1_1 - T^2_2\right) - 3 W_{(s)}\right] \nonumber \\
& + &
\frac{e^{(\lambda - \nu)/2} r^3}{4} \left(\ddot\lambda +
\frac{\dot\lambda^2}{2} - \frac{\dot\lambda \dot\nu}{2}\right)
\label{pre}
\end{eqnarray}

\noindent
which can be formally integrated to give

\begin{eqnarray}
m_T = C r^3
& + & r^3 \int^r_0{\frac{e^{(\nu + \lambda)/2}}{r^4}
\left[4 \pi r^3 \left(T^1_1 - T^2_2\right) - 3 W_{(s)}\right] dr}
 \nonumber \\
& + &
r^3 \int^r_0{\frac{e^{(\lambda - \nu)/2}}{4r}\left(\ddot\lambda +
\frac{\dot\lambda^2}{2} - \frac{\dot\lambda \dot\nu}{2}\right) dr}
\label{forin}
\end{eqnarray}

\noindent
or equivalently

\begin{eqnarray}
m_T = C r^3
& + & r^3 \int^{r_\Sigma}_0{\frac{e^{(\nu + \lambda)/2}}{r^4}
\left[4 \pi r^3 \left(T^1_1 - T^2_2\right) - 3 W_{(s)}\right] dr}
\nonumber \\
& + &
r^3 \int^{r_\Sigma}_0{\frac{e^{(\lambda - \nu)/2}}{4r}\left(\ddot\lambda +
\frac{\dot\lambda^2}{2} - \frac{\dot\lambda \dot\nu}{2}\right) dr}
\nonumber \\
& - & r^3 \int^{r_\Sigma}_r{\frac{e^{(\nu + \lambda)/2}}{r^4}
\left[4 \pi r^3 \left(T^1_1 - T^2_2\right) - 3 W_{(s)}\right] dr}
\nonumber \\
& - &
r^3 \int^{r_\Sigma}_r{\frac{e^{(\lambda - \nu)/2}}{4r}\left(\ddot\lambda +
\frac{\dot\lambda^2}{2} - \frac{\dot\lambda \dot\nu}{2}\right) dr}
\label{or}
\end{eqnarray}

\noindent
Finally, evaluating (\ref{forin}) at $r_\Sigma$ to obtain $C$,
and replacing it in (\ref{or}), we obtain, using (\ref{Ws})
and (\ref{Wint}),

\begin{eqnarray}
m_T & = & (m_T)_\Sigma \left(\frac{r}{r_\Sigma}\right)^3 \nonumber \\
& - & r^3 \int^{r_\Sigma}_r{e^{(\nu+\lambda)/2} \left[\frac{8 \pi}{r}
\left(T^1_1 - T^2_2\right)
+ \frac{1}{r^4} \int^r_0{4 \pi \tilde{r}^3 (T^0_0)' d\tilde{r}}
 \right] dr} \nonumber \\
& - & r^3 \int^{r_\Sigma}_r{
\frac{e^{(\lambda-\nu)/2}}{2r}\left(\ddot\lambda + \frac{\dot\lambda^2}{2}
- \frac{\dot\lambda \dot\nu}{2}\right)
  dr}
\label{emte}
\end{eqnarray}

\noindent
In the static (or quasi-static) case
($\ddot\lambda = \dot\lambda^2 = \dot\lambda \dot\nu = 0$)
the expression above is identical to eq.(32) in \cite{HeSa}.

\noindent
Let us now assume that our system, initially at rest and in equilibrium,
is perturbed and departs from equilibrium. Since $\omega$ is the fluid
velocity as
 measured by our local Minkowski observer,
then immediately after
this departure we have

\begin{equation}
\omega \approx 0 \qquad \; \qquad \dot\omega \not= 0
\label{ia}
\end{equation}

\noindent
and, from (\ref{feq01}) and (\ref{T01}),

\begin{equation}
\dot\lambda \approx 0
\label{th}
\end{equation}

\noindent
On the other hand, the following expression may be easily
obtained for $\ddot\lambda$ from (\ref{feq01}) and (\ref{T01}):

\begin{eqnarray}
\ddot\lambda & = & -  8 \pi r e^{(\nu + \lambda)/2}
[\,
\left(\rho + P_r\right) \frac{\omega}{1-\omega^2}
\frac{\dot\nu}{2}
 +
\frac{\left(\rho+P_r\right) \omega}{1-\omega^2}
\frac{\dot\lambda}{2} \nonumber \\
& + &
\left(\dot\rho + \dot P_r\right)
\frac{\omega}{1-\omega^2}
+
\left(\rho+P_r\right) \dot\omega
\frac{1+\omega^2}{\left(1-\omega^2\right)^{2}}
\,]
\label{ddl}
\end{eqnarray}

\noindent
or, evaluating this immediately after the departure from equilibrium,

\begin{equation}
\ddot\lambda = - 8 \pi r e^{(\nu + \lambda)/2}
\left[\left(\rho + P_r\right) \dot\omega\right]
\label{ddla}
\end{equation}

\noindent
Inserting (\ref{th}) and (\ref{ddla}) into (\ref{emte}), we finally obtain,
using (\ref{T00}), (\ref{T11}) and (\ref{T2233})

\begin{eqnarray}
m_T & = & (m_T)_\Sigma \left(\frac{r}{r_\Sigma}\right)^3 \nonumber \\
& + & 4 \pi r^3 \int^{r_\Sigma}_r{e^{(\nu + \lambda)/2}
\left[\frac{2}{r} \left(P_r - P_\bot\right) -
\frac{1}{r^4} \int^r_0{\tilde{r}^3 \rho' d\tilde{r}}\right] dr} \nonumber \\
& + & 4 \pi r^3 \int^{r_\Sigma}_r
{e^{\lambda} \left(\rho + P_r\right) \dot\omega dr}
\label{mfi}
\end{eqnarray}

\noindent
Note
that boundary conditions

\begin{equation}
m_\Sigma = M \qquad  , \qquad   \nu_\Sigma = - \lambda_\Sigma \qquad  , \qquad
\left[P_r\right]_\Sigma = 0
\label{bou}
\end{equation}

\noindent
where $M$ is the total mass (the mass parameter in the Scwarzschild metric),
 imply from (\ref{I+II}) and (\ref{T11})

\begin{equation}
(m_T)_\Sigma = M + \left(\frac{\rho \omega^2}{1 - \omega^2}\right)_\Sigma
\label{imp}
\end{equation}

\noindent
Thus, although $(m_T)_\Sigma = M$ in the static (and quasi-static)
case and immediately after the departure from equilibrium, this is no
longer true in general in the dynamic case.We shall discuss on
eq.(\ref{mfi}) in the next section.

\section{Discussion}

\noindent
Let us asssume that our system consists of pure incoherent dust
($P_r = P_\bot = 0$). Then the value of $m_T$ for a spherical
region of radius $r$ within $\Sigma$, immediately after the
system departs from equilibrium, depends on three different
contributions.
The first one, $M \left(r/r_\Sigma\right)^3$, is the gravitational
mass of a static, homogeneous sphere, of radius $r$, within $\Sigma$.
The second contribution,
$
- 4 \pi r^3 \int^{r_\Sigma}_r{\left(e^{(\nu+\lambda)/2}/r^{4}\right)
\left(\int^r_0{\tilde r^3 \rho' d\tilde r}\right) dr} ,
$
depends on the inhomogeneity of the
energy-density distribution and will be positive if,
as required by stability conditions, $\rho'<0$.
The last term
$
4 \pi r^3 \int^{r_\Sigma}_r{e^{\lambda}
\left(\rho + P_r\right) \dot\omega dr}
\label{lt}
$
will contribute negatively in the case of
collapse ($\dot\omega < 0$).
Thus, density inhomogeneity tends to increase
the Tolman mass within the sphere of radius $r$, if
$\rho' < 0$ , leading thereby (according to (\ref{a}))
to a faster collapse.
 This conclusion
is also true in the case of slowly collapsing
spheres, since (\ref{mfi}) is identical to (\ref{13To}) in \cite{HeSa}
except for the last term containing $\dot\omega$.
This last term, in turn, tends to ``stabilize'' the
system by decreasing (increasing) the Tolman mass
in the process of collapse (expansion).

\noindent
In the case of an anisotropic fluid, the term
$
4 \pi r^3 \int^{r_\Sigma}_r{e^{(\nu + \lambda)/2}
\left(2/r\right) \left(P_r - P_\bot\right) dr}
$
plays a similar role to the density inhomogeneity, with
$P_r > P_\bot$ ($P_\bot > P_r$) corresponding to
$\rho' < 0 $ ($\rho' > 0$).
The fact that $P_r > P_\bot$ ($P_\bot > P_r$) leads to
a stronger (weaker) collapse is, intuitively, in agreement
with eq.(\ref{Prp}).
This conclusion concerning the influence
of local anisotropy is also valid in the slowly evolving regime
\cite{HeSa}.
However, in this case,  we only have information
about the system at the  moment it departs
from equilibrium. A complete description of the evolution
of the system, requires a full integration of the field equations.

\noindent
We have seen so far why and how the pace of collapse
is affected by the density inhomogeneity and local anisotropy.
Since the former is related to the formation of naked
singularities, it could be speculated that local anisotropy
might also lead to the formation of naked singularities.
Of course, only a specific example could confirm (or discard)
this suspicion.



\begin{thebibliography}{}

\bibitem[1]{Var}
Eardley D. M. and Smarr L., 1979, Phys. Rev. D {\bf 19}, 2239;
Christodoulou D., 1984, Commun. Math. Phys. {\bf 93}, 171;
Newman R. P. A. C., 1986, Class. Quantum Grav {\bf 3}, 527;
Waugh B. and Lake K., 1988, Phys. Rev. D {\bf 38},1315;
Dwivedi I. H. and Joshi P. S., 1992, Class. Quantum Grav {\bf 9}, L69;
Joshi P. S. and Dwivedi I. H., 1993, Phys. Rev. D {\bf 47}, 5357;
Singh T. P. and Joshi P. S., 1996, Class. Quantum Grav {\bf 13}, 559.

\bibitem[2]{Opp}
Oppenheimer J and Snyder H., 1939, Phys. Rev. {\bf 56}, 455.

\bibitem[3]{Bon}
Bondi H., 1964, Proc. R. Soc. London, {\bf A281}, 39.

\bibitem[4]{BowLia}
Bowers R. and Liang E., 1974, Astrophys. J. {\bf 188}, 657.

\bibitem[5]{MisSh}
Misner C. and Sharp D., 1964, Phys. Rev. {\bf 136}, B571.

\bibitem[6]{Cah}
Cahill M. and McVittie G., 1970, J. Math. Phys. {\bf 11}, 1382.

\bibitem[7]{HeSa}
Herrera L.and Santos N. O., 1995, Gen. Rel. Grav., {\bf 27}, 1071.

\bibitem[8]{To}
Tolman R., 1930, Phys. Rev., {\bf 35}, 875.

\bibitem[9]{Coo}
Cooperstock F. I., Sarracino R. S. and Bayin S. S.,
1981, J. Phys. A {\bf 14}, 181.

\bibitem[10]{Deu}
Devitt J. and Florides P. S., 1989, Gen. Rel. Grav. {\bf 21}, 585.

\bibitem[11]{Gro}
Gr{\o}n {\O}., 1985, Phys. Rev. D, {\bf 31}, 2129.

\bibitem[12]{Lig}
Lightman A., Press W., Price R. and Teukolsky S., 1975,
Problem Book in Relativity and Gravitation (Princeton University Press,
Princeton).

\end{thebibliography}
\end{document}